\documentclass[prb,aps]{revtex4}
\usepackage{graphicx}

\begin{document}

\title{Thermal spin current and magnetothermopower by Seebeck spin tunneling}
\vspace*{-9mm}
\author{R. Jansen$^1$, A.M. Deac$^2$, H. Saito$^1$ and S. Yuasa$^1$}
\affiliation{$^1\,$National Institute of Advanced Industrial Science and Technology (AIST),
Spintronics Research Center, Tsukuba, Ibaraki, 305-8568, Japan.\\
$^2\,$Helmholtz-Zentrum Dresden-Rossendorf, Institute of Ion Beam Physics and Materials Research,
01314 Dresden, Germany.}

\begin{abstract}
The recently observed Seebeck spin tunneling, the thermoelectric analog of spin-polarized
tunneling, is described. The fundamental origin is the spin dependence of the Seebeck coefficient
of a tunnel junction with at least one ferromagnetic electrode. Seebeck spin tunneling creates a
thermal flow of spin-angular momentum across a tunnel barrier without a charge tunnel current. In
ferromagnet/insulator/semiconductor tunnel junctions this can be used to induce a spin accumulation
$\Delta\mu$ in the semiconductor in response to a temperature difference $\Delta T$ between the
electrodes. A phenomenological framework is presented to describe the thermal spin transport in
terms of parameters that can be obtained from experiment or theory. Key ingredients are a
spin-polarized thermoelectric tunnel conductance and a tunnel spin polarization with non-zero
energy derivative, resulting in different Seebeck tunnel coefficients $S_{st}^{\,\uparrow}$ and
$S_{st}^{\,\downarrow}$ for majority and minority spin electrons. We evaluate the thermal spin
current, the induced spin accumulation and $\Delta\mu/\Delta T$, discuss limiting regimes, and
compare thermal and electrical flow of spin across a tunnel barrier. A salient feature is that the
thermally-induced spin accumulation is maximal for smaller tunnel resistance, in contrast to the
electrically-induced spin accumulation that suffers from the impedance mismatch between a
ferromagnetic metal and a semiconductor. The thermally-induced spin accumulation produces an
additional thermovoltage proportional to $\Delta\mu$, which can significantly enhance the
conventional charge thermopower. Owing to the Hanle effect, the thermopower can also be manipulated
with a magnetic field, producing a Hanle magnetothermopower.
\end{abstract}

\maketitle

\vspace*{-7mm}

\section{Introduction}
\indent The interplay of heat and charge transport is the basis of thermoelectrics, enabling the
conversion of heat flow to electrical power, and vice-versa. Spintronics concerns the interplay of
spin and charge transport and has transformed magnetic data storage technology and magnetic field
sensing. The connection between these two important fields has been established in studies of
thermoelectric properties of magnetic nanostructures
\cite{johnson,wang,falko1,falko2,fukushima1,fukushima2,ansermet1,ansermet2,hatamitorque,hatami}.
This interplay between heat and spin transport, now referred to as spin-caloritronics
\cite{hatamitorque,bauer1}, has recently gained impetus because the combination of thermoelectrics
and spintronics offers unique possibilities. On the one hand, it provides a new, spin-based
approach to thermoelectric power generation and cooling. On the other hand, it provides a thermal
route to create and control the flow of spin in novel spintronic devices that make functional use
of heat and temperature gradients. In addition, most spintronic nanodevices involve the application
of electrical currents, which create thermal gradients that might influence magnetic and
spin-related phenomena and thereby device performance and efficiency. This underpins
the importance of understanding the fundamental interactions between thermal and spin effects.\\
\indent A notable recent development is the observation of the spin Seebeck effect by Uchida et al.
\cite{uchida1}. They found that when a ferromagnetic material (permalloy) is subjected to a thermal
gradient $\nabla T$, a spin current is injected into a non-magnetic metal (Pt) strip attached to
the ferromagnet. This spin current is converted into a voltage proportional to $\nabla T$ via the
inverse spin Hall effect \cite{uchida1}. The name "spin Seebeck effect" suggests it is the spin
analogue of the classical charge Seebeck effect. The latter can be understood in the following way,
noting that the electrical conductance depends on the energy of the charge carriers. A thermal
gradient across a (non-magnetic) conductor causes a flow of electrons with energy above the Fermi
energy from the hot to the cold side. Simultaneously, electrons with energy below the Fermi energy
flow in the opposite direction. There is a net current because the two current components do not
cancel when the conductivity for electrons above and below the Fermi energy is different. In open
circuit conditions this results in a voltage between the hot and cold end of the conductor,
proportional to $S\,\nabla T$, with $S$ the Seebeck coefficient. In a ferromagnetic conductor one
expects that the Seebeck coefficient is different for electrons with majority and minority spin, as
their electronic properties are different. A thermal gradient across a ferromagnet would then yield
a net flow of spin parallel to the thermal gradient, and produce in a spin voltage (accumulation of
spin) at the hot and the cold ends of the ferromagnet. Although this was originally suggested to be
the cause of the observed spin Seebeck effect \cite{uchida1}, the currently accepted interpretation
is rather different. It is now considered to originate from a non-equilibrium between the magnon
distribution in the ferromagnet and the electrons in the attached non-magnetic metal, resulting in
thermally driven dynamical spin pumping across the interface, without a global spin current or spin
accumulation in the ferromagnet \cite{bauer2,sinova,adachi1,adachi2,uchidareview}. This microscopic
mechanism bears no relation with the classical charge Seebeck effect. Yet, the spin Seebeck effect
is a novel method to convert a thermal gradient into a voltage, via the spin, and the phenomenon is
generic, i.e., subsequent to the original demonstration for permalloy \cite{uchida1}, it was also
observed in ferromagnetic insulators \cite{uchida2}, ferromagnetic semiconductors
\cite{myers1,myers2} and other ferromagnetic metals \cite{takanashi}.\\
\indent In a different type of experiment, Slachter et al. \cite{slachter} demonstrated for the
first time that the Seebeck coefficient of a ferromagnet depends on spin. They showed that a
thermal gradient in ferromagnetic permalloy induces a spin current in the permalloy parallel to the
heat flow, and that when the heat flow is directed towards an interface with a non-magnetic metal,
the spin current crosses the interface and produces a spin accumulation in the non-magnetic metal.
This thermally-driven spin injection is directly proportional to the difference of the Seebeck
coefficient of majority and minority spin electrons {\em in} the ferromagnet, which was shown to be
a fraction of the regular charge Seebeck coefficient of the ferromagnet \cite{slachter,slachtermodel}.\\
\indent A distinctly different phenomenon, Seebeck spin tunneling, was observed by Le Breton et al.
\cite{lebreton}. Unlike previous work, Seebeck spin tunneling is a pure interface effect that
occurs in tunnel junctions with a temperature difference $\Delta T$ between the two electrodes,
provided that at least one of the electrodes is ferromagnetic. It was demonstrated that Seebeck
spin tunneling creates a flow of spin angular momentum across a tunnel barrier {\em without} a
charge tunnel current. This thermal spin current was shown to be governed by the variation of the
spin polarization of the tunneling process with the energy of the tunneling electrons. As will be
explained here, Seebeck spin tunneling is directly linked to the spin-dependent Seebeck coefficient
of a magnetic tunnel contact. This implies that the results of Le Breton et al. effectively
demonstrated that the Seebeck tunnel coefficient for majority and minority spin is different. In
addition, Le Breton et al. {\em used} Seebeck spin tunneling for thermally-driven spin injection
into a semiconductor, i.e., they observed that in ferromagnet/insulator/silicon tunnel contacts,
the thermal spin current induces a spin accumulation
$\Delta\mu$ in the silicon.\\
\indent An interesting analogy exists between electrical and thermal spin transport across a tunnel
junction. Seebeck spin tunneling is the thermoelectric analogue of spin-polarized tunneling, which
refers to the spin dependence of the electrical conductance of a magnetic tunnel contact. The
latter was clearly demonstrated four decades ago in experiments \cite{meservey1,meservey2} on
ferromagnet/insulator/superconductor junctions, showing that the charge tunnel current between a
ferromagnet and a non-magnetic counter electrode is spin polarized. In magnetic tunnel junctions
comprising {\em two} ferromagnetic electrodes, spin-polarized tunneling also gives rise to large
tunnel magnetoresistance \cite{moodera,yuasa}, denoting the change of the tunnel resistance as a
function of the relative orientation (parallel vs. antiparallel) of the magnetization of the
electrodes. Analogously, Seebeck spin tunneling produces a tunnel magnetothermopower, i.e., a
dependence of the thermopower of a magnetic tunnel junction on the relative magnetization alignment
of the two electrodes. This tunnel magnetothermopower (or tunnel magneto-Seebeck effect) has been
theoretically predicted \cite{wang,heiliger} and recently observed by different groups, first in
MgO-based tunnel junctions \cite{munzenberg,liebing}, and subsequently in Al$_2$O$_3$ junctions
\cite{mangin}. Anisotropy of the tunnel magnetothermopower was also reported \cite{molenkamp}. Last
but not least, it was predicted that thermal gradients give rise to thermal spin-transfer torques
in magnetic heterostructures \cite{hatamitorque,slonczewski} and tunnel junctions \cite{xia} and
experimental evidence for thermal torques has been presented \cite{ansermet3,rezende}.\\
\indent Le Breton et al. \cite{lebreton} described the salient features of Seebeck spin tunneling
by numerical evaluation of a free electron model. Here we present a phenomenological framework to
describe Seebeck spin tunneling in linear response in terms of parameters that can be obtained from
experiment and analytical or {\em ab-initio} theory. Key ingredients are a tunnel conductance with
a spin polarization that depends on energy, and the spin polarization of the thermally-induced
electrical transport across the tunnel barrier. An important aim is to establish the connection
with a Seebeck tunnel coefficient that depends on spin. We evaluate the thermal spin current, the
induced spin accumulation and $\Delta\mu/\Delta T$, and show that these are proportional to
$S_{st}^{\,\uparrow}-S_{st}^{\,\downarrow}$, where $S_{st}^{\,\uparrow}$ and
$S_{st}^{\,\downarrow}$ denote the Seebeck tunnel coefficient for majority and minority spin,
respectively. We discuss limiting regimes, and point out that the thermally-induced spin
accumulation increases for smaller tunnel resistance, in contrast to the electrically-induced spin
accumulation that suffers from the impedance mismatch between a ferromagnetic metal and a
semiconductor \cite{schmidt,rashba,fertprb}. We also compare the fundamental limits of thermal and
electrical spin tunneling. Finally, we demonstrate that the thermally-induced spin accumulation
produces an additional thermovoltage proportional to $\Delta\mu$ that can significantly enhance the
conventional charge thermopower. The thermopower can be manipulated with a magnetic field owing to
the Hanle effect, producing a Hanle magnetothermopower in junctions with only one ferromagnetic
electrode.

\section{Seebeck spin tunneling}
\subsection{Model}
\indent We consider a tunnel junction with a ferromagnetic electrode and a non-ferromagnetic
electrode, typically a semiconductor (Fig. 1) or metal. It is assumed that the tunnel resistance
$R_{tun}$ is much larger than the resistance of the electrodes, such that tunneling limits the
transport. The ferromagnetic and non-magnetic electrode are characterized by so-called spin
resistances $r_s^{fm}$ and $r_{s}$, respectively, describing the ratio of the spin accumulation in
the material and the associated spin current due to spin relaxation
\cite{fertprb,fertIEEE,jaffres}. The value of $R_{tun}$ relative to $r_s^{fm}$ and $r_{s}$ plays an
important role in the spin transport, as it determines the coupling between the two spin systems.
We will assume that $R_{tun}>>r_s^{fm}$, which is usually the case when an interface with a
(Schottky or oxide) tunnel barrier is formed (the resistance-times-area product is 1-10 $\Omega\mu
m^2$ or larger for magnetic tunnel junctions \cite{yuasa,tunnelRA}, while $r_s^{fm}$ is typically
below 0.01 $\Omega\mu m^2$ for transition metal ferromagnets \cite{fertprb,fertIEEE,jaffres}).
Therefore, one does not need to consider the spin accumulation and spin-dependent (electrical or
thermal) transport {\em within} the ferromagnet, including the spin-dependent Seebeck effect
\cite{slachter,slachtermodel} due to any temperature gradients within the ferromagnet. The spin
resistance of the non-magnetic material is normally much larger than $r_s^{fm}$. We do not make any
specific assumptions about the value of $R_{tun}$ relative to $r_{s}$. We thus cover the regime
with $R_{tun}>r_{s}$ where the spin accumulation in the semiconductor is effectively decoupled from
the ferromagnet by the tunnel barrier, as well as the regime with $R_{tun}<r_{s}$ where the
coupling to the ferromagnet reduces the spin accumulation in the non-magnetic material
\cite{schmidt,rashba,fertprb}.

\begin{figure}[htb]
\hspace*{0mm}\includegraphics*[width=80mm]{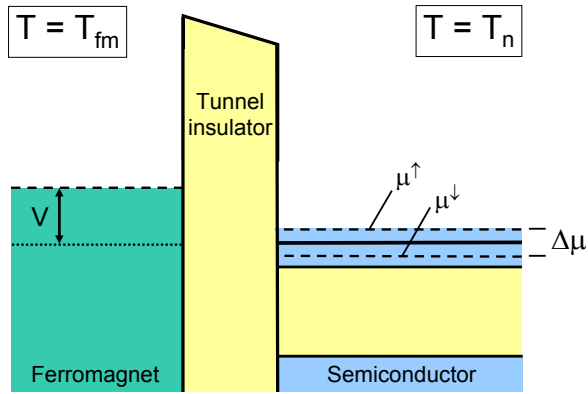} \caption{Energy band diagram of a
ferromagnet/insulator/semiconductor tunnel junction. The semiconductor electrode is at temperature
$T_{n}$, whereas the ferromagnet is at $T_{fm}$. A spin accumulation exists in the semiconductor,
described by a spin splitting $\Delta\mu = \mu^{\uparrow}-\mu^{\downarrow}$ of the electrochemical
potential. The applied bias voltage $V$ is referenced to the spin-average potentials.} \label{fig1}
\end{figure}

\indent For a given bias voltage $V$ and temperature difference $\Delta T$ across the barrier, the
tunnel current $I^{\sigma}$ for each spin ($\sigma = \,\uparrow,\downarrow$, denoting electrons
with magnetic moment, respectively, parallel and anti-parallel to the magnetization of the
ferromagnet) is:
\begin{eqnarray}
I^{\uparrow} = G^{\uparrow}\left(V-\frac{\Delta\mu}{2}\right) + L^{\uparrow} \Delta T \label{eq1}\\
I^{\downarrow} = G^{\downarrow}\left(V+\frac{\Delta\mu}{2}\right) + L^{\downarrow} \Delta T
\label{eq2}
\end{eqnarray}
The first term on the right-hand side describes the electrically driven current governed by
spin-dependent tunnel conductances $G^{\uparrow}$ and $G^{\downarrow}$. It incorporates the effect
of the shifts of the electrochemical potentials $\mu^{\sigma}$ in the semiconductor due to the
presence of the spin accumulation $\Delta\mu = \mu^{\uparrow}-\mu^{\downarrow}$ (for convenience we
have defined $\Delta\mu$ in units of volt). Note that the spin accumulation typically decays
exponentially with distance from the injection contact and that $\Delta\mu$ denotes the value of
the spin accumulation at the interface with the tunnel barrier, as is relevant for tunneling. The
second term on the right-hand side describes the thermally-induced tunnel current in response to a
temperature difference, as governed by $L^{\uparrow}$ and $L^{\downarrow}$, which we will refer to
as the thermoelectric tunnel conductances (not to be confused with the thermal conductance that
describes heat flow). We define $\Delta T=T_{n}-T_{fm}$, where $T_{n}$ and $T_{fm}$ are the
temperatures of the non-magnetic and ferromagnetic electrode, respectively, and $V=V_{n}-V_{fm}$,
where $V_{n}$ and $V_{fm}$ are the spin-averaged potentials of the non-magnetic and ferromagnetic
electrode.\\
\indent The total conductances are $G = G^{\uparrow}+G^{\downarrow}$ and $L =
L^{\uparrow}+L^{\downarrow}$, and their spin polarizations are $P_G =
(G^{\uparrow}-G^{\downarrow})/(G^{\uparrow}+G^{\downarrow})$ and $P_L =
(L^{\uparrow}-L^{\downarrow})/(L^{\uparrow}+L^{\downarrow})$. The charge tunnel current $I$ and the
spin tunnel current $I_s$ are then:
\begin{eqnarray}
I = I^{\uparrow}+I^{\downarrow} = G\,V - P_G\,G\,\left(\frac{\Delta\mu}{2}\right) + L \,\Delta T \label{eq3} \\
I_s = I^{\uparrow}-I^{\downarrow} = P_G\,G\,V - G \left(\frac{\Delta\mu}{2}\right) +
P_L\,L\,\Delta T \label{eq4}
\end{eqnarray}
The spin current consists of an electrical ($P_G\,G\,V$) and a thermal contribution
($P_L\,L\,\Delta T$), as well as a correction due to the $\Delta\mu$ that is induced by the
(electrical and/or thermal) spin current. The feedback of $\Delta\mu$ on the spin/charge tunnel
current implies that another (independent) relation between $\Delta\mu$ and $I_s$ is required to
obtain a solution. This is provided by the requirement of a steady-state spin accumulation in the
non-magnetic material, which implies that the spin current $I_s$ injected by tunneling is balanced
by the spin current due to spin relaxation in the material, integrated over its full spatial
extent. The spin current associated with spin relaxation is proportional to the spin accumulation.
We define a spin resistance $r_s$ of the non-magnetic material via:
\begin{equation}
\Delta\mu = 2\,I_s\,r_s \label{eq5}
\end{equation}
In our model, $r_s$ is a phenomenological parameter that describes the conversion of the spin
current $I_s$ that is injected by tunneling, into a spin accumulation $\Delta\mu$. As mentioned
before, $\Delta\mu$ denotes the value of the spin accumulation right at the tunnel interface. This
definition of $r_s$ makes no specific assumptions about the spatial profile of the spin
accumulation in the non-magnetic material, or the formalism used to compute it. If we use the
spin-diffusion equation and a spin accumulation that decays exponentially with distance from the
injection interface with the spin-diffusion length $L_{sd}$, then the spin resistance of a unit
contact area can be expressed as $\rho_n L_{sd}$, where $\rho_n$ is the resistivity of the
non-magnetic material \cite{fertprb,fertIEEE,jaffres}. This result is frequently used to analyze
experimental data, but note that it requires introduction of a somewhat unusual factor of two in
eqn. (\ref{eq5}).

\subsection{Spin current and Seebeck spin tunnel coefficient}
\indent Equations (\ref{eq3}-\ref{eq5}) fully define the system and allow us to obtain the relevant
quantities. We first derive a general expression for the spin accumulation, valid for electrical
($I\neq0$) and thermal ($\Delta T\neq0$) injection, as well as for a combination of the two. We
shall discuss the case of purely thermal ($I=0$) and purely electrical ($\Delta T=0$) driving force
later on. The solutions for $\Delta\mu$ and the spin current in terms of $\Delta T$ and $I$ are:
\begin{equation}
\Delta\mu = \left\{\frac{2\,r_s}{R_{tun} + (1 - P_G^2)\,r_s}\right\}\left[\,(P_G)\,R_{tun}\,I -(P_L
- P_G)\,S_0\Delta T\,\right] \label{eq8}
\end{equation}
\begin{equation}
I_s = \left\{\frac{1}{R_{tun} + (1 - P_G^2)\,r_s}\right\}\left[\,(P_G)\,R_{tun}\,I -(P_L -
P_G)\,S_0\Delta T\,\right] \label{eq9}
\end{equation}
where $R_{tun}=1/G$ is the tunnel resistance and $S_0=-L/G$ is the charge thermopower (in the
absence of a spin accumulation; see below). The first term in eqn. (\ref{eq9}), proportional to
$I$, is the electrical spin current associated with the spin-polarized charge current. The second
term is the pure spin current due to Seebeck spin tunneling (driven by $\Delta T$) and will be
referred to as the Seebeck spin current. The Seebeck spin tunneling coefficient $S_{st}=\Delta\mu /
\Delta T$ is obtained by setting $I=0$ in eqn. (\ref{eq8}), and rewriting $(P_L - P_G)\,S_0$ in
terms of spin-dependent Seebeck tunnel coefficients $S_{st}^{\uparrow}=-L^{\uparrow}/G^{\uparrow}$
and $S_{st}^{\downarrow}=-L^{\downarrow}/G^{\downarrow}$ for majority and minority spin,
respectively. We then obtain an important result, namely, $\Delta\mu / \Delta T$ is proportional to
$(S_{st}^{\uparrow} - S_{st}^{\downarrow})$:
\begin{equation}
S_{st} = \frac{\Delta\mu}{\Delta T} = \left\{\frac{(1 - P_G^2)\,r_s}{R_{tun} + (1 -
P_G^2)\,r_s}\right\}(S_{st}^{\uparrow} - S_{st}^{\downarrow}) \label{eq11c}
\end{equation}
Since $0 \leq P_G^2 \leq 1$, the pre-factor always has a positive sign. The sign of $S_{st}$ is
thus determined by the difference between $S_{st}^{\uparrow}$ and $S_{st}^{\downarrow}$. Note that
the pre-factor tends to zero when the tunnel spin polarization becomes very large ($P_G \approx 1$
or $-1$), but $S_{st}$ does not because also $G^{\sigma}$ for one of the two spin channels goes to
zero. Hence, either $S_{st}^{\uparrow}$ or $S_{st}^{\downarrow}$ diverges. Taking this into
account, one finds that $S_{st} = +4\,(L^{\downarrow}/G^{\uparrow})(r_s/R_{tun})$ for $P_G=1$ and
$S_{st} = -4\,(L^{\uparrow}/G^{\downarrow})(r_s/R_{tun})$ for $P_G=-1$. Expressions (\ref{eq8}) and
(\ref{eq9}) apply to situations for which $I$ and $\Delta T$ are fixed. This includes recent
experiments on Seebeck spin tunneling \cite{lebreton} (where $I=0$), as well as most experiments on
electrical spin injection that are performed in constant current mode (and $\Delta T=0$).
Alternatively, we can express $\Delta\mu$ and $I_s$ in terms of $\Delta T$ and $V$. However, care
has to be taken not to set $V=I\,R_{tun}$, as this is not correct when $\Delta\mu\neq0$ or $\Delta
T\neq0$, see eqn. (\ref{eq3}).\\

\subsection{Charge thermopower and Hanle magnetothermopower}
\indent The charge thermopower S is obtained from the voltage $V|_{I=0}$ for which $I$ vanishes.
From eqn. (\ref{eq3}) we obtain:
\begin{equation}
S = \frac{V|_{I=0}}{\Delta T} = S_0 + \left(\frac{P_G}{2}\right) S_{st} \label{eq13}
\end{equation}
This is another important result. In the presence of a spin accumulation, the charge thermopower is
not equal to $S_0=-L/G$. There is an additional, previously unidentified, contribution that is
proportional to $\Delta\mu$ and thus to the Seebeck spin tunnel coefficient $S_{st}$. Since $P_G$
can be positive or negative depending on the properties of the ferromagnet/insulator interface, and
also $S_{st}$ can have either sign, the additional contribution can enhance or reduce the charge
thermopower. The enhancement can be significant because $S_{st}$ can be much larger than $S_0$, as
we will see in the discussion section.\\
\indent Next we address how the thermally-induced spin accumulation can be detected as a voltage
signal. Just as for electrically-induced spin accumulation, this can be done via the Hanle effect,
which occurs when the spins in the semiconductor are subjected to a magnetic field $B$ at a solid
angle $\theta$ with the spin direction \cite{opticalorientation,dash,invertedhanle}. This causes
spin precession and consequently a reduction of $\Delta\mu$ depending on $\theta$ and on the
product of the spin lifetime $\tau_s$ and the Larmor frequency $\omega_L = g\mu_BB / \hbar$, where
$g$ is the Land\'e g-factor, ${\mu}_{B}$ the Bohr magneton and $\hbar$ Planck's constant divided by
2$\pi$. When the tunnel resistance is sufficiently large ($R_{tun} \gg r_s$) such that the coupling
of the spin accumulation to the ferromagnet can be neglected, spin precession causes a decay of
$\Delta\mu$ in a Lorentzian fashion \cite{opticalorientation,invertedhanle}:
\begin{equation}
\Delta\mu = 2\,I_s\,r_s \equiv 2\,I_s\,r_{s}^{0}\,\left\{cos^2(\theta) + \frac{sin^2(\theta)}{1 +
(\omega_L \tau_s)^2}\right\} \label{eq19}
\end{equation}
In the absence of any magnetic field there is no spin precession and the spin resistance is
$r_{s}^{0}$. If we keep $\Delta T$ and the charge current $I$ constant, apply a magnetic field
perpendicular to the spins, and increase $B$ from zero to a value for which $\omega_L\tau_s\gg1$,
the spin resistance is gradually reduced from $r_{s}^{0}$ to zero. The $\Delta\mu$ then also goes
to zero, even if the spin current $I_s$ that is injected by tunneling is non-zero. This results in
the desired voltage change, which is obtained from Eqn. (\ref{eq3}) and (\ref{eq19}) as:
\begin{equation}
\Delta V_{Hanle} = V|_{\omega_L=0} - V|_{\omega_L\tau_s\gg1} =
\left(\frac{P_G}{2}\right)\Delta\mu|_{\omega_L=0} \label{eq20}
\end{equation}
An important point is that this expression is valid irrespective of how the spin accumulation is
created. In other words, also for a thermally-induced spin accumulation, the detected voltage
signal $\Delta V_{Hanle}$ is given by $P_G/2$ times $\Delta \mu$, which is the same relation as for
electrically-induced spin accumulation \cite{dash}.\\
\indent In the regime where $R_{tun}<r_s$, the magnitude of $\Delta\mu$ is reduced by the coupling
of the spins to the ferromagnet, but also the functional dependence of $\Delta\mu$ on $B$ is
modified and eqn. (\ref{eq19}) does not correctly describe the dependence on $B$ and $\theta$.
However, the maximum and minimum values of $\Delta\mu$ for, respectively, $\omega_L=0$ and
$\omega_L\tau_s\gg1$ are still properly described. Therefore, the amplitude of the spin
accumulation can still be correctly obtained from  eqn. (\ref{eq20}). However, extracting the spin
lifetime from the line width of the Hanle curve requires a detailed
description of the line shape taking the interaction with the ferromagnet into account.\\
\indent The ability to manipulate the spin accumulation with an external magnetic field (owing to
the Hanle effect) also means that $S_{st}$ and hence the charge thermopower $S$ can be controlled
by a magnetic field. We define the Hanle magnetothermopower $S_{mag}$ as the relative change of the
thermopower between its value in zero magnetic field, and the value at $\Delta\mu=0$ that
corresponds to $\omega_L\tau_s\gg1$:
\begin{equation}
S_{mag} = \frac{S|_{\omega_L=0} - S|_{\omega_L\tau_s\gg1}}{S|_{\omega_L\tau_s\gg1}} =
\left(\frac{P_G}{2}\right)\,\left(\frac{S_{st}|_{\omega_L=0}}{S_0}\right) \label{eq24}
\end{equation}
The magnetothermopower is mediated by the spin accumulation, has a variation with magnetic field
that is governed by the Hanle effect, and occurs in tunnel contacts in which only one of the
electrodes is ferromagnetic. It is thus different from the tunnel magnetothermopower recently
observed in a magnetic tunnel junction with two ferromagnetic electrodes, which has a magnetic
field variation that is controlled by the angle between the magnetizations of the two electrodes
\cite{munzenberg,liebing}. The Hanle magnetothermopower can be very large because $S_{st}$ can be
much larger than $S_0$.

\section{Discussion}
\subsection{Origin and definition of Seebeck spin tunneling}
\indent We have seen that Seebeck spin tunneling occurs when the Seebeck coefficient of a magnetic
tunnel contact is different for majority and minority spin. Le Breton et al. \cite{lebreton}
described the salient features of Seebeck spin tunneling by numerical evaluation of a free electron
model, and showed that Seebeck spin tunneling is determined by the energy derivative of the tunnel
spin polarization. In this section we will establish the important connection between these two
notions, and also clarify the definition of Seebeck spin tunneling.\\
\indent First, we note that $(S_{st}^{\uparrow} - S_{st}^{\downarrow})$ is proportional to $(P_L -
P_G)$ and that the spin accumulation induced by Seebeck spin tunneling is proportional to $(P_L -
P_G)$. This is easily understood because when $I=0$, any thermally-induced current (with
polarization $P_L$) must be balanced by an equal but opposite electrically-driven current (with
polarization $P_G$). If the tunnel spin polarization does {\em not depend on energy}, all the
induced (electrical or thermal) current components necessarily have the same spin polarization, and
we have $P_L = P_G$. In that case $S_{st}=0$, the Seebeck spin current vanishes for all $\Delta T$,
and a spin current exists only if the charge current is non-zero. To illustrate that $P_L = P_G$ if
the tunnel spin polarization does not depend on energy, we express $G^{\sigma}$ and $L^{\sigma}$ in
terms of the tunnel transmission function $D^{\sigma}(E)$ integrated over energy $E$, as
\cite{sivan,heiliger}:
\begin{eqnarray}
G^{\sigma} = -\frac{e^2}{h} \int D^{\sigma}(E)\,(\partial_{E}f(E,\mu,T))\,dE \label{eq11a}\\
L^{\sigma} = -\frac{e}{h}\frac{1}{T} \int D^{\sigma}(E)\,(E-\mu)\,(\partial_{E}f(E,\mu,T))\,dE
\label{eq11b}
\end{eqnarray}
where $\partial_{E}f(E,\mu,T)$ is the energy derivative of the Fermi-Dirac distribution function
$f(E,\mu,T)$. When $D^{\uparrow}(E)$ and $D^{\downarrow}(E)$ have the same variation with energy,
we can write $D^{\sigma}(E)=\chi^{\sigma}D(E)$, where the coefficients $\chi^{\uparrow}$ and
$\chi^{\downarrow}$ do not depend on energy. Inserting this in eqns. (\ref{eq11a}) and
(\ref{eq11b}) we find $P_G = P_L =
(\chi^{\uparrow}-\chi^{\downarrow})/(\chi^{\uparrow}+\chi^{\downarrow})$.
Then $P_G$ and $P_L$ are independent of $E$ and $P_L = P_G$. Since $(S_{st}^{\uparrow} -
S_{st}^{\downarrow})$ is proportional to $(P_L - P_G)$, we conclude that
$S_{st}^{\uparrow}=S_{st}^{\downarrow}$ if the tunnel spin polarization does not depend on energy.
This establishes the connection between the energy derivative of $P_G$ and a spin-dependent Seebeck
coefficient: $S_{st}^{\uparrow} \neq S_{st}^{\downarrow}$ only if the tunnel spin polarization
depends on energy.\\
\indent There is ample evidence for the energy dependence of the tunnel spin polarization. Indirect
evidence comes from the decay of tunnel magnetoresistance with bias voltage in magnetic tunnel
junctions \cite{moodera,yuasa}. Direct evidence is provided in two reports for transition metal
ferromagnets on Al$_2$O$_3$, where the variation of the tunnel spin polarization with energy of the
tunnel electrons was determined \cite{valenzuela,park}. A significant asymmetry in the decay of the
tunnel spin polarization with energy below and above the Fermi energy was reported, the decay
being much faster above the Fermi energy.\\
\indent Let us define the term Seebeck spin tunneling more precisely, because one could argue that
a thermally-driven spin current can exist even if $P_L = P_G$. While technically correct, in this
case the charge current is non-zero, i.e., it is not a pure spin current. In fact, for $P_L = P_G$
any thermally-induced spin current is from the spin-polarized charge current that arises from the
shift of the $I-V$ curve by an amount equal to the charge thermovoltage $S_0\Delta T$. We do not
consider this to be Seebeck spin tunneling, which is associated with a non-zero energy derivative
of the tunnel spin polarization (in analogy with the conventional charge Seebeck effect that is
related to a non-zero energy derivative of the charge conductivity). Experimentally, one will have
a combination of a thermally-driven spin-polarized charge current and Seebeck spin
tunneling, unless one measures at $I=0$, as done in Ref. \onlinecite{lebreton}.\\

\subsection{Comparison of electrical and thermal spin current}
\indent It is instructive to compare the magnitude of the spin current due to electrical and
thermal spin tunneling. Besides the fundamental interest, this is of course important from a
technological point of view. One question is whether the creation of a spin current by a
temperature difference across the tunnel barrier can be more energy efficient than creating a spin
current electrically. Another question is whether the heat that is produced by electrical
generation of a spin current can be re-used to supplement it with a thermal spin current, and how
much increase in spin current, or reduction in energy consumption, can be obtained in this way. The
answer to those questions cannot be given in general terms. The efficiency of creating the
temperature difference depends crucially on the thermal design of the structures. Moreover, whereas
it has been known for four decades that the electrical tunnel conductance is spin polarized and the
polarization has been rather well optimized, the Seebeck spin tunneling has only recently been
observed and the difference $(S_{st}^{\uparrow} - S_{st}^{\downarrow})$ is far from optimum. We
will therefore only
discuss the factors that determine the ultimate limits of electrical and thermal spin current.\\
\indent We consider the driving term as well as the proportionality factor between $\Delta\mu$ and
the driving term (see Table 1). The thermal driving term is $S_0 \Delta T$, which should be
compared to the electrical driving term $R_{tun}\,I$. For non-magnetic metal tunnel junctions $S_0$
has been evaluated \cite{leavens,marschall} to be in the range of $50-100\,\mu$V/K, although it has
been predicted that it can be enhanced by magnons in ferromagnetic tunnel junctions
\cite{falko1,falko2}. The $\Delta T$ for tunnel junctions will, in practice, be limited to about 10
K. Hence, $S_0 \Delta T$ is of the order of 1 mV, which is to be compared to typical values of a
few 100 mV for $R_{tun}\,I$. In general, the thermal driving term will thus be smaller than the
electrical driving term.\\
\indent With respect to the proportionality factor, for electrical spin injection it is limited by
$P_G$, since its absolute value cannot be larger than 1 (by definition). However, such a
restriction does not exist for the proportionality factor of the thermal spin current, since there
is no limit for the energy derivatives that govern $(S_{st}^{\uparrow} - S_{st}^{\downarrow})$ and
$(P_L-P_G)$. In  principle, $L^{\uparrow}$ and $L^{\downarrow}$ can be equal in magnitude but of
opposite sign, so that $L\approx0$ and $P_L$ goes to infinity. Physically, this corresponds to the
situation where the tunnel spin polarizations for states above and below the Fermi energy have
opposite sign, such that one type of spin is driven from the hot to the cold side of the tunnel
contact, and the other type of spin is driven from the cold to the hot side. Hence, the
proportionality factor for thermal spin accumulation can, in principle, be arbitrarily large for
suitably engineered materials. This can therefore (more than) compensate for the smaller thermal
driving term. This suggests that Seebeck spin tunneling can be a viable approach to create a spin
current, either by itself, or in conjunction with an electrical spin current.
\begin{table}[h]
\caption{Comparison of thermal and electrical spin current in a tunnel junction.} \label{table}
\vspace*{2mm}
\begin{tabular}{ c c | c c c | c c c c c | c c c c c }
\\
Method & & & Type of spin current & & & Driving term & & Typical values & & & Polarization factor & & Extreme values & \\
\hline \vspace*{-0mm} & & & & & & & & & & & & & &\\
Electrical & & & Spin-polarized charge current & & & $R_{tun}\,I$ & & $\sim 100$ mV & & & $P_G$ & & $\pm \,1$ & \\
\vspace*{-2mm} & & & & & & & & & & & & & &\\
Thermal & & & Pure spin current (I=0) & & & $S_0 \Delta T$ & & $\sim 0.1-1$ mV & & & $S_{st}^{\uparrow}-S_{st}^{\downarrow}$ & & $\pm\,\infty$ & \\
\vspace*{-0mm} & & & & & & & & & & & & & &\\
\hline
\end{tabular}
\end{table}

\subsection{Magnitude of Seebeck spin tunnel coefficient}
\indent The magnitude of the thermal spin current (and spin accumulation) depends on the value of
the polarizations $P_L$ and $P_G$, as well as on the coupling of the spin accumulation to the
ferromagnet. An important point is that the Seebeck spin tunnel coefficient $S_{st}$ can be much
larger than the regular charge thermopower $S_0$. To illustrate this, the ratio of $S_{st}$ and
$S_0$ is shown as a function of relevant parameters in Fig. 2. For materials with large tunnel spin
polarization ($P_G \approx 1$ or $-1$), a very large Seebeck spin tunneling coefficient is produced
if $P_L$ and $P_G$ are unequal, or preferably, of opposite sign. This situation would occur for
ferromagnet/insulator interfaces that have an almost full spin polarization of the tunnel
conductance for states at and below the Fermi energy $E_F$, but a rapidly decaying or even opposite
spin polarization above $E_F$, for instance due to the onset of a contribution to the tunneling of
a band with opposite spin orientation. Since the total thermopower is given by the sum of $S_0$ and
$(P_G/2)\,S_{st}$ (eqn. (\ref{eq13})), the thermally-induced spin accumulation in the non-magnetic
material can significantly enhance the charge thermopower of a tunnel junction.

\begin{figure}[htb]
\includegraphics*[width=80mm]{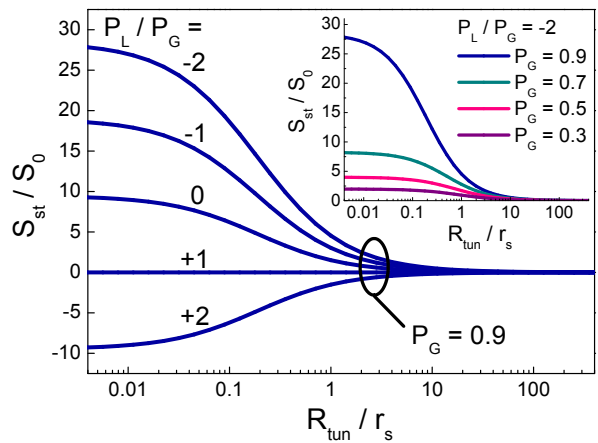}
\vspace*{-3mm}\caption{Seebeck spin tunnel coefficient $S_{st}$ normalized to $S_0$, as a function
of $R_{tun}/r_s$, for various values of $P_L/P_G$ and fixed $P_G=0.9$, and (inset) for fixed
$P_L/P_G=-2$ and $P_G$ varied. Note that $S_{st}=0$ if $P_L/P_G=+1$.} \label{fig2}
\end{figure}

\subsection{Scaling with tunnel resistance}
\indent A noteworthy difference between electrical and thermal creation of a spin accumulation is
the scaling with tunnel resistance (Fig. 3, and appendix A for explicit expressions for the
different regimes). For electrical spin injection, the polarization of the injected current
($I_s/I$) is $P_G$ as long as the tunnel resistance is larger than the spin resistance of the
semiconductor (see appendix A, eqn. (\ref{eq18})). However, when $R_{tun} \ll r_s$, the coupling to
the ferromagnet starts to play a role, and the feedback of $\Delta\mu$ on the tunnel transport
severely reduces the spin polarization of the tunnel current (which is well established
\cite{schmidt,rashba,fertprb}). As a result, $\Delta\mu/I$, the spin accumulation per unit injected
charge current, is constant at large $R_{tun}$ but decays at small tunnel resistance (Fig. 3, top panel).\\
\indent In contrast, the scaling of the thermally-induced spin accumulation is opposite: $\Delta\mu
/ \Delta T$ {\em increases} as the tunnel resistance is lowered, and reaches a large and constant
value when $R_{tun}$ becomes {\em smaller} than $r_s$ (fig. 3, bottom panel). This behavior is
consistent with that obtained by numerical evaluation of a free electron model \cite{lebreton}. It
thus appears that thermal injection is more efficient at small tunnel resistance, whereas
electrical creation of spin accumulation requires sufficiently large tunnel resistance to overcome
the impedance mismatch. Note that in Fig. 3 we have neglected that $S_0$ decays for thinner tunnel
barriers, because the decay is known to be very weak \cite{leavens,marschall} and does not
critically affect the main scaling trend. It is also known that for an ultrathin tunnel contact,
the thermal (heat) conductance $I_Q/\Delta T$, with $I_Q$ the heat current, is limited by the
interfaces rather than the bulk thermal heat conductance of the tunnel barrier material. Therefore,
the thermal conductance $I_Q/\Delta T$ is expected to be approximately independent of $R_{tun}$. A
similar trend would thus result if we would plot $\Delta\mu/I_Q$ instead of $\Delta\mu / \Delta T$.
We thus find that the spin accumulation per unit {\em charge} current is maximum for {\em large}
tunnel resistance, whereas the spin accumulation per unit {\em heat} current across the tunnel
barrier is maximum at {\em small} tunnel resistance.

\begin{figure}[htb]
\includegraphics*[width=80mm]{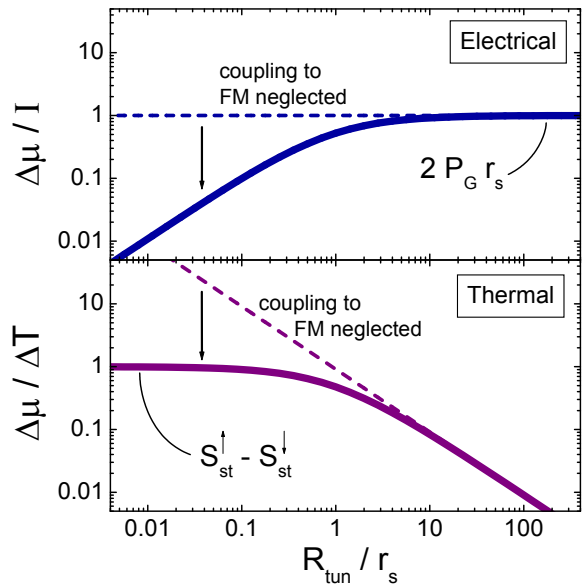}
\vspace*{-3mm}\caption{Scaling of the electrical and thermal spin accumulation with tunnel
resistance. Shown are $\Delta\mu/I$ for electrical injection (top panel) and $\Delta\mu / \Delta T$
for Seebeck spin tunneling (bottom panel), both as a function of the ratio $R_{tun}/r_s$ of the
tunnel resistance and the spin resistance of the semiconductor. The results are normalized to the
maximum value as indicated ($2\,P_G\,r_s$ for electrical and
$S_{st}^{\uparrow}-S_{st}^{\downarrow}$ for thermal). The dashed lines describe the result when one
neglects the feedback of $\Delta\mu$ on the spin current injected from the ferromagnet. The arrows
indicate the reduction due to the feedback. The $P_G$ was set to 0.3.} \label{fig3}
\end{figure}

\section{Summary}
\indent A phenomenological framework has been presented to describe Seebeck spin tunneling, the
thermoelectric analog of spin-polarized tunneling. It was established that Seebeck spin tunneling
originates from the spin dependence of the Seebeck coefficient of a tunnel junction with a
ferromagnetic electrode, i.e., $S_{st}^{\,\uparrow} \neq S_{st}^{\,\downarrow}$. The connection
with a tunnel spin polarization $P_G$ that depends on energy was also made. Seebeck spin tunneling
creates a thermal flow of spin-angular momentum across a tunnel barrier without a charge tunnel
current. In ferromagnet/insulator/semiconductor tunnel junctions it allows creation of a spin
accumulation $\Delta\mu$ in the semiconductor by a temperature difference $\Delta T$ between the
electrodes. We expressed the thermal spin current, the induced spin accumulation and
$\Delta\mu/\Delta T$ in terms of the spin-dependent Seebeck coefficients, tunnel resistance and
spin resistance of the non-magnetic electrode. The thermally-induced spin accumulation produces an
additional thermovoltage proportional to $\Delta\mu$, which can significantly enhance the
conventional charge thermopower. Because the spin accumulation can be manipulated via the Hanle
effect, the thermopower depends on a magnetic field, producing a Hanle magnetothermopower in
junctions in which only one of the electrodes is a ferromagnet. The thermally-induced spin
accumulation was shown to be maximum for smaller tunnel resistance, in contrast to the
electrically-induced spin accumulation that suffers from the impedance mismatch between a
ferromagnetic metal and a semiconductor. While the efficiency of electrical spin injection is
limited by the fact that $|P_G|\leq1$, no such restriction exists for thermal spin current that is
determined by the energy derivative of $P_G$, which is unbounded. With suitably engineered
materials, Seebeck spin tunneling is thus a viable option for efficient creation of spin current.


\begin{appendix}
\section{Spin current and accumulation in limiting regimes}

\subsection{Thermal spin current}
\indent There are two limiting regimes for Seebeck spin tunneling ($I=0$). When $R_{tun} \gg r_s$,
the induced $\Delta\mu$ remains relatively small and the feedback of $\Delta\mu$ on the tunnel
transport is negligible. For this regime one obtains:
\begin{equation}
\frac{\Delta\mu}{\Delta T} \approx -2\,\left(P_L - P_G\right)\left(\frac{r_s}{R_{tun}}\right)S_0 =
(1 - P_G^2)\left(\frac{r_s}{R_{tun}}\right)(S_{st}^{\uparrow} - S_{st}^{\downarrow}) \label{eq14}
\end{equation}
The spin accumulation decays at larger tunnel resistance since for a tunnel contact $S_0 = -L/G$
depends only weakly on $R_{tun}$. This is because all (thermal or electrical) tunnel current
components, and hence $L$ and $G$, decay exponentially with tunnel barrier width and height
\cite{leavens,marschall}.\\
When $R_{tun}/r_s<(1 - P_G^2)$, we have:
\begin{equation}
\frac{\Delta\mu}{\Delta T} \approx \frac{-2\left(P_L - P_G\right)}{1 - P_G^2}\,S_0 =
(S_{st}^{\uparrow} - S_{st}^{\downarrow}) \label{eq15}
\end{equation}
In this regime, which corresponds to tunnel contacts with sufficiently low resistance-area product,
$S_{st}$ and $\Delta\mu$ do not directly depend on $R_{tun}$.\\

\subsection{Electrical spin current}
\indent For comparison, the spin accumulation $\Delta\mu^{el}$ induced by electrical injection of a
spin-polarized charge current, without a temperature difference across the tunnel barrier, is given
by:
\begin{equation}
\Delta\mu^{el} = \frac{2\,r_s\,R_{tun}}{R_{tun} + (1 - P_G^2)\,r_s}\,P_G\,I \label{eq17}
\end{equation}
and the spin current is:
\begin{equation}
I_s^{el} = \frac{R_{tun}}{R_{tun} + (1 - P_G^2)\,r_s}\,P_G\,I \label{eq18}
\end{equation}
We remark that it is customary \cite{fertprb,fertIEEE,jaffres,rashba} to replace the real tunnel
resistance $R_{tun}$ by $(1 - P_G^2)\,r^{*}_B$, introducing $r^{*}_B$ as an effective tunnel
resistance. The pre-factor then takes a more simple form without the factor $(1 - P_G^2)$. As a
result, it is no longer evident that in order to determine the transition into the regime where the
feedback of $\Delta\mu$ on the tunneling becomes relevant, one has to compare the tunnel resistance
to $(1 - P_G^2)\,r_s$. The transition thus depends on the
value of $P_G$. We therefore choose to retain the term $(1 - P_G^2)$ explicitly.\\

\end{appendix}

\end{document}